\documentclass[prd]{revtex4}
\usepackage{graphicx}
\usepackage{color}

\usepackage{latexsym}
\usepackage{hyperref}

\begin{document}

\bigskip

\title{ A CONSISTENT SCENARIO FOR $B\to PS$ DECAYS }
\vskip 6ex
\author{D. Delepine}
\email{delepine@fisica.ugto.mx}
\author{J.\ L.\ Lucio M.}
\email{lucio@fisica.ugto.mx}
\affiliation{Instituto de F\'{\i}sica, Universidad de Guanajuato \\
Loma del Bosque \# 103, Lomas del Campestre, \\
37150 Le\'on, Guanajuato; M\'exico}
\author{J. A. Mendoza S.}
\email{jairoam@unipamplona.edu.co}
\affiliation{Depto. de Física-Matem\'aticas, Universidad de Pamplona \\
Pamplona, Norte de Santander, Colombia.}
\author{Carlos A. Ram\'{\i}rez}
\email{jpjdramirez@yahoo.com}
\affiliation{Escuela de F\'{\i}sica, Universidad Industrial de Santander, \\
A.A. 678, Bucaramanga, Colombia }

\begin{abstract}
\noindent We consider  $B\to PS$  decays where $P$ stands for pseudoscalar
and $S$ for a heavy (~1500 MeV) scalar meson. We achieve agreement with
available experimental data -- which includes a two orders of magnitude hierarchy --
assuming the scalars mesons are two quark states. The contribution of the dipolar
penguin operator ${\cal O}_{11}$ is quantified.
\end{abstract}
\maketitle

\section{INTRODUCTION}

\noindent The scalar sector below two GeV is poorly understood,
nevertheless several features --like the presence of two
multiplets and several of their properties -- naturally arise in
the analysis of a  number of authors. A first set of  scalars
with masses around 1.5 GeV \cite{scalars} are grouped
in a heavy multiplet,
including the $K_0^*(1430)$, $a_0(1450)$, $f_0(1500)$ for the
octet, $f_0(1370)$ which is identified with the singlet and the
$f_0(1710)$ which seems to be mainly glueball. The octet is nearly
degenerate, like similar pseudoscalar, vector, axial vector and tensor
multiplets, their widths are small ($\leq 100$ MeV). The mixing angles seems to be small except by the singlet-glueball which is around $-20^{\rm o}$, according to H. Y. Cheng in ref. \cite{chengm}. It has been more difficult to establish
the lighter multiplet,  even the existence and
nature of some of their members is in doubt. The light multiplet should
include the $a_0(980)$, $f_0(980)$ and the $\kappa=K_0^*(800)$ in
the octet; while the singlet could be identified with the
$\sigma=f_0(600)$. The mixing is not clear and their widths are
very large.  Ideally, the former multiplet can be identified as the
ground state of quark antiquark bound states with angular momenta
one while the later with the ground state of four quarks systems
with zero angular momenta. In the real world an undetermined
mixing between the two multiplets is expected. Alternatively both
multiplets could be identified as quark-antiquark states with
angular momenta one, the lighter being the ground state while the
heavier  the first excited state.
\bigskip

\noindent The full understanding of the scalar multiplets previously
described remain a  challenge, both from the experimental perspective
as well as from the theoretical  point of view \cite{scalars}. To start with,
there is not enough and conclusive experimental information regarding
the existence and properties of the scalars. Notice that the  information
is  poor not because of the  lack of sources of scalar mesons, for example
many of the decays of  particles containing $c$ or $b$ quarks involves
the production of scalar mesons. The information on the scalars is scarce
because of the large width they have since that produces a large
overlap with nearby resonances and with the background. In spite
of those problems, precise experimental results are available \cite{scalars,pdg,f0}
for the mass and width of the $f_0$ and $K_0^*$, for  the $\beta$ angle
 \cite{beta} of the CKM matrix and for several partial widths.
It has been speculated that
the $\alpha$ angle can be extracted in processes involving scalars
\cite{laplace} and new projects like the LHCB \cite{newprojects}
will improve the old measurements and obtain new results.
Relevant to our work are the branching ratios for the $B \to PS$
decays measured by different groups, which show a non trivial
hierarchy. The experimental data collected in Table I suggest that, for $B\to PS$
decays including members of the  heavy scalar multiplet, the order of magnitude of the
 branching ratios  involving the $K_0^*(1430)$, the $f_0(1370), f_0(1500)$
and the $a_0(1450)$ are different.

\bigskip

\noindent On the theoretical side the situation is not better. The origin of the difficulties are
the  non perturbative regime of QCD and the  limited  computer capacity
for the lattice approach. The nature  of the observed
scalars has been discussed at length and proposals exist to
identify them as 2 or 4 quark states, glueballs, molecules,
etc. and  several theoretical formalisms have been
developed to calculate non leptonic decays. The simplest one is
the so called \lq Naive Factorization Approach' (NFA) \cite{nfa},
which in general produces the correct order of magnitude and its
predictions are in rough agreement with the experimental results.
Discrepancies are known to occur in two cases, for \lq color
suppressed' processes and when important re-scattering effects are involved,
for example processes where direct CP violation is relevant \cite{nfacri,fsi}.
The advantage of formalism where a
systematic expansion is implemented and where higher order
correction can be organized and controlled are of great importance
(QCDF, SCET, pQCD, LCSR, etc. \cite{lcsr,pqcd,qcdf,scet}), in
particular when high accuracy predictions are required.
\bigskip

\noindent Additional reasons to study the
$B \to PS$ decays are:  they offer a window to study the spectroscopy
and the dynamics of the scalar sector, the $B \to 3P$ decays get a contribution
from the $B \to PS,\ PV,\ PT$, so that  in order to achieve an
appropriated estimate for the former decay the latter must be well
known  \cite{bennich}. In a similar way one can argue that in
order to extract signals of possible new physics, the contribution
of low lying conventional physics has to be known in detail,
including the contributions of the scalar mesons \cite{giri}.  We believe
that the understanding of  the physical origin of the hierarchy of scales appearing
in  the $B \to PS$ decays can shed some light on the nature of the scalars \cite{Delepine,delepine1}. Complementary information on the nature of the scalars  may be obtained from $D\to PS$ physics \cite{dtoscalars}: in the first case through the decay constants, $\bar f^s$ while in the latter through the $F^{DS}$ form factors.
The purpose of the present work is to consider the $B \to PS$ decays with $S$ a member of the
heavy scalar multiplet. We assume that the leading contribution to these processes
is given by the NFA and that, in first approximation, contributions other than the
leading one can be safely neglected. In these conditions  the dominant contribution
can be clearly identified and  the existence of the scales in the branching ratios naturally
arises. Besides the NFA our approach can be
summarized along the following lines: we include ten dimension six
four quark operators and the dimension five chromomagnetic
operator $O_{11}$ \cite{O11}, annihilation contributions are
included and the form factors required are obtained by using sum
rules, so infrared divergences are absent. This approach, together with
 $SU(3)$ symmetry, allows us to reproduce the pattern observed experimentally.

\section{BRANCHING RATIOS AND MIXING}

\noindent Our results are summarized in Table I.  It is worth remarking that both the experimental  data and our results points to the existence of branching ratios that ranges from 45 to  0.5 (in units of $10^{-6}$).  In the following paragraphs
we introduce the notation, conventions and explain the procedure we follow to
obtain these branching ratios. Within the NFA  the hadronic matrix
elements can be reduced to products of decay constants and form
factors. In order to achieve this one  uses the \lq vacuum
saturation' approximation and neglect other  intermediate states. This seems to be
a reasonable assumption since the hadronic resonances have masses  in the $1-2$ GeV range,
far from the $m_b$ region. For the invariant amplitude we write
${\cal M}_{f\to i} =   ~ <f|H|i> ~    = ~   G_FA_{f\to i}/\sqrt{2}$  while the branching ratios are given
 by  $B=\tau_BG_F^2|A|^2p/16\pi m_B^2=\tau_BG_F^2|A|^2/32\pi m_B$, with
$\tau_B$ the $B$ lifetime. The decay constants and form factors
are defined as \cite{Delepine,delepine1,nfa}:

\begin{eqnarray}
\langle P(p)|A_\mu|0 \rangle &=& - i f_P p_\mu ; \hskip1cm \langle
S(p) | V_{\mu}| 0 \rangle =f_Sp_\mu = {m_2-m_1\over m_S} \bar f_S p_\mu \hskip1cm \langle f_0 | q \bar q| 0 \rangle = m_{f_0} \bar f_{f_0}, \nonumber \\
\langle S(p_2)|L_\mu|P(p_1)\rangle &=&-i
\left[\left(p_1+p_2-{m_1^2-m_2^2\over q^2}q\right)_\mu
F_+^{M_1M_2} +{m_1^2-m_2^2\over q^2}q_\mu F_0^{M_1M_2}(q^2)\right]
\end{eqnarray}

\noindent with $q=p_1-p_2$.

\begin{table}
\begin{center}
\begin{tabular}{|l|ccc|c|c|c|c|c|} \hline
Decay  &  BELLE   & BABAR  &  HFAG \cite{pdg}& $B_{\rm exp.}$  & NFA &
NFA+${\cal O}_{11}$ & QCDF \cite{Delepine} & pQCD \cite{Delepine}  \\ \hline

$\pi^-a_0^+(1450)(\pi\eta)$
&    &  $<2.3^*$ & $<2.3^*$ &&   8 &&3.1 &  \\

$\pi^+a_0^-(1450)$
& $ $   & &&&    2 &&0.5 & \\

$\pi^-a_0^0(1450)$
& $ $   &  & &&  4 &&2.5 &    \\

$\pi^-f_0(1370)$
& & $<3$  &$<3$  &  & && &  \\

$\pi^-f_0(1500)$
& &  & && &&&   \\

$\pi^0a_0^-(1450)$ & $ $   &  & &&   0.01&&1.1 &   \\ \hline

$K^+a_0^-(1450)$
& $<3.1^*$   & $ $ &$<3.1^*$ &&1 &&0.3 &   \\

$K^+a_0^0(1450)$
& $ $   &  & &&0.5 &&0.2 &   \\

$K^-f_0(1370)(\pi\pi)$
& & $<10.7^*$  &$<10.7^*$  & $<41$  &8 &7 & &  \\

$K^-f_0(1500)(\pi\pi)$
&                       &$0.73\pm 0.21\pm 0.47^*$ & $0.7(5)^*$   & 2(1)  &23 &21 & & 55 \\

$\bar K^0a_0^-(1450)$
& $ $    &  &  &&&&0.1 &   \\

$K^0a_0^0(1450)$
& $ $    &  & &&&&0.1 &   \\

$K^0f_0(1370)$
&                 &      &       &     &7  &7 &   &  \\

$K^0f_0(1500)$
 &       &     &                               &  &22 &21 & &42  \\  \hline

$\pi^-K_0^{*+}( K^+\pi^0)$
&$49.7\pm 3.8\pm 3.8^{+1.2}_{-4.8}$     & $25.4^{+3.0+6.1}_{-3.7-5.6}$    &34(5)  &34(5)  &45     &45         &11 &43  \\

$\pi^+K_0^{*0}( K^+\pi^-)$
&$51.6\pm 1.7\pm 6.8^{+1.8}_{-3.1}$     &$32.2\pm 1.2^{10.8}_{-6}$           &45(6)  &45(6)  &45 (in) &45 (in)   &11 &48  \\

$\pi^0 K_0^*{}^+$
 &  &  & & &25 &25 &5.3&29\\

$\eta K_0^*{}^+$
&$^{}_{}$ & $15.8\pm 2.2\pm 1.4\pm 1.7 $           &16(3)  &16(3) &7 &7 &&\\

$\pi^0K_0^*{}^0$
&           & $11.7^{+1.4+4}_{-1.3-3.6}$ &12(4)&12(4) &17 &17 &6.4 &18\\

$\eta K_0^*{}^0$ &$^{}_{}$ & $9.6\pm 1.4\pm 0.7\pm 1.1 $ &10(2)
&10(2) &7 &7 &&\\ \hline

\end{tabular}
\label{tab1}
 \caption{Branching ratios for the $ B \to PS$ decays
 (in units of $10^{-6}$), for the heavier scalar multiplet.
 The values reported for the widths marked with $^*$ include the corresponding branching of the scalar decaying channel. To obtain the NFA predictions we used $B(f_0(1370)\to 2\pi)=0.26(1)$,
 $B(f_0(1500)\to 2\pi)=0.35(2)$ and  for the $a_0(1450)\to \pi \eta$ no reliable value exists\cite{pdg}.}
\end{center}
\end{table}
\bigskip

\noindent We have left to the appendix details regarding
the effective Hamiltonian we use - which  includes ten dimension six operators and the so called  ${\cal O}_{11}$ operator - and the matrix elements evaluation. The most interesting decays are those involving the
$S=K_0^*(1430)$ both because they have the largest branching ratio
(around 40, in units of $10^{-6}$) and  because the
theoretical predictions are the cleanest.  The $a_6$  term is  by far the dominant
one. The amplitudes are proportional to
$\lambda_{ts}f_{K_0^*}a_6m_{K_0^*}^2/m_sm_b\sim  \lambda_{ts}a_6m_b m_{K_0^*} \bar f_{K_0^*}$ times $SU(3)$ factors. The origin of the
enhancement is a combination of large CKM matrix elements, the
novanishing decay constant and a large $m_{K_0^*}$ (Chiral enhancement) mass. The $SU(3)$
symmetry allow us to relate different decays involving the $K_0^*$
and so, by  measuring one of them, one can predict the others, a fact
that is not distorted by the ${\cal O}_{11}$ contributions. For
the numerical analysis we used the following input parameters:
$F^{B\pi}=0.27(4)$, $F^{BK}=0.33(4)$, $m_s(2.1)=90$ MeV,
 $F^{Ba_0(1450)}=F^{BK_0^*(1430)}=0.26$ and, when required, $SU(3)$ relations are invoked.
Although predictions for  $f_{K_0^*}$  are available
\cite{Delepine},  we prefered to
include  the $B^+\to \pi^+K_0^{*0}$   experimental value as an
input, obtaining thus  $f_{K_0^*}^{\rm eff.} \simeq 58 $ MeV
($f_{K_0^*}^{\rm eff.} \simeq 56 $ MeV when the ${\cal O}_{11}$ is
taken into account). The  branching ratios we obtain for  other channels involving the
$K_0^*$ are reported in Table I. Notice that the value obtained for  $f_{K_0^*}$ is not far from
the theoretical predictions (see Table II).
\bigskip

\noindent We now consider the decays involving
$S=f_0(1370),\ f_0(1500)$ and $f_0(1700)$. Their relevance stem
from the  large branching ratios predicted for them  \cite{Delepine} -- of
the same order as the $K_0^*$  -- and also due to the possible
glueball nature of the  $f_0(1700)$.  Their amplitudes are proportional to
$\lambda_{ts}a_6m_b m_{K_0^*} \bar f_{f_0}^s$ times $SU(3)$ factors and mixing angles ($s$- quark content).
Our predictions for these processes are included in Table I, unfortunately the
experimental results are still inconclusive. Note that except
the $f_{0}(1500)$  decay channel,  the NFA plus  $SU(3)$ symmetry
for the heavy scalar multiplet leads predictions for  the
 branching ratios in rough agreement with the
experimental values. However,  even if the experimental data is
poor the discrepancy between our results and experimental data
is evident, there is  a one order of magnitude difference. In this sense it is important to
remark that in order to obtain the results of tabler I we assumed, following H. Y. Cheng \cite{chengm} a mixing
between the  glueball, singlet and
octet components given by:

\begin{eqnarray}
\left(\begin{array}{c}f_0(1370)\\ f_0(1500)\\ f_0(1700)
\end{array} \right)&=&\left(\begin{array}{ccc} 0.78 & 0.51 & -0.36 \\ -0.54 & 0.84 & 0.03\\ 0.32 & 0.18 & 0.93
\end{array} \right)\left(\begin{array}{c}N\\ S\\ G
\end{array} \right)  \nonumber \\
&=& \left(\begin{array}{ccc}
c_{12}c_{13}&s_{12}c_{13}&s_{13}\\
-s_{12}c_{23}-c_{12}s_{23}s_{13}&c_{12}c_{23}-s_{12}s_{23}s_{13}& s_{23}c_{13}\\
s_{12}s_{23}-c_{12}c_{23}s_{13}&-s_{23}c_{12}-s_{12}c_{23}s_{13}&c_{23}c_{13}
\end{array}\right)\left( \begin{array}{ccc} \sqrt{2\over 3} & \sqrt{1\over 3}& 0\\ -\sqrt{1\over 3}& \sqrt{2\over 3}& 0\\ 0& 0& 1 \end{array} \right)\left(\begin{array}{c}N\\ S\\ G
\end{array} \right) \nonumber \\
\end{eqnarray}

\noindent where $s_i=\sin \theta_i$ and so on. The angles $\theta_{12}\simeq 2^{\rm o}$, $\theta_{13}\simeq -21^{\rm o}$ and $\theta_{23}\simeq 2^{\rm o}$ are the mixing between  singlet-octet, singlet-glueball and octet-glueball, respectively. The singlet and the octet are $f_0(1370)\sim f_{\rm sing.}=\sqrt{2/3}\ N+S/\sqrt{3}$, $f_0(1500)\sim f_{\rm oct.}=N/\sqrt{3}-S\sqrt{2/3}$,  $S=\bar ss$, $N=\left(\bar uu+\bar dd\right)/\sqrt{2}$ and $G=gg$ the glueball. Thus, in this approach \cite{chengm},    there is only an small mixing between the singlet and the glueball. Using these values the prediction for $B\to f_0(1500)K$ is in conflict with the experimental data.
One way to avoid this problem is to leave  $\theta_{12}$ as a free parameter,  keeping the others fixed. Using the experimental data
we obtain the following inequality  for  the mixing between the singlet and the octet:

\begin{eqnarray}
|-s_{12}\sqrt{\frac{1}{3}}+c_{12}\sqrt{\frac{2}{3}}|&\leq& 0.34.
\end{eqnarray}

\noindent  These constraints lead two possible values:

\begin{eqnarray}
35^{\circ} \leq &\theta_{12}& \leq 74^{\circ} \\
215^{\circ} \leq &\theta_{12}&\leq 254^{\circ}
\end{eqnarray}

\noindent It is worth noticing that these values for the mixing are close to those mentioned by several groups  \cite{scalars}.
\bigskip

\noindent Finally for the decays involving the  $a_0 ( 1450 )$,
$B\to a_0(1450)\pi, a_0(1450)K$, the terms proportional to $a_4-a_6\sim 0$ almost vanish and the branching ratios are smaller.
Two different cases must be considered.  The first when the
amplitude is dominated by the tree level contribution $a_1$ (The amplitudes are proportional to  $\lambda_{ud}a_1m_B^2 f_\pi $), then
the theoretical prediction is reliable and the branchings are
predicted  to be are around $10$ (in units of $10^{-6}$). The
second case arises when no tree level contribution exist and terms like annihilation are dominant. In this
case the branchings are of order 0.1-1 (in units of $10^{-6}$) but
the theoretical uncertainties are larger since other contributions
(FSI for example\cite{fsi}) maybe important. Unfortunately little
is known about these corrections.

\begin{table}
\begin{center}
\begin{tabular}{|l||l|l|l|l|} \hline

Ref.        & $(f/\bar f)_{K_0^*(1430)}, \bar f$  &$\bar
f_{a_0(1450)}$ & $\bar f_{f_0(1500)}^s$ & $m_s$ [GeV] \\ \hline

Meurice-87 \cite{maltman} & 27 &-         &-      & \\ \hline

Narison-89 \cite{maltman} &40(6)          & -     &-      & \\
\hline

Maltman \cite{maltman} &42(2)      &390(159)      &-      &\\
\hline

Chernyak-01 \cite{maltman} &70(10)     & -     &-      &\\ \hline

Shakin-01 \cite{maltman} & 30  & 207         &-      & \\ \hline

Pennington-01 \cite{maltman} &-         &-          &-      & \\
\hline

Du-04 \cite{maltman} &42(8),\ 427(85) &      & -     & 0.14
\\ \hline

Cheng-05 \cite{Delepine} at $\mu=1$ GeV &445(50)  &460(50)
&490(50)&0.119     \\ \hline

Cheng-05 \cite{Delepine} at $\mu=2.1$ GeV &550(60)  &570(60)
&605(60)&0.09 \\ \hline

lattice-06 \cite{scalars} &&&&  \\ \hline

\end{tabular}
\end{center}

\caption{Decay constants for scalars (in MeV). The heavy scalars
are assumed to be two quark states. Notice that the
constants computed by Cheng, were obtained by using sum rules, OPE
and Renormalization Group equations that render $\bar f$ scale
dependant.}
\end{table}
\bigskip

\section{SUMMARY}

\noindent In this work we studied the $B \rightarrow PS$ decay
where $S$ stands for a member of the heavy scalar multiplet. The
computation have been done assuming  the heavy scalar multiplet is
a two quark states, using  $SU(3)$  symmetry and  the naive
factorization approach. Our conclusions can be summarized as
follows:

\begin{itemize}

\item Within the error bars, it is possible to reproduce the hierarchy of branching ratios experimentally observed in the  $B \to PS$ decays, whether or not  the operator ${\cal O}_{11}$ is included.

\item When the singlet-octet mixing given by \cite{scalars} is used,  we obtain a prediction for the $f_0(1500)$ which is   one order of magnitude above the experimental limit. A solution to this problem can be obtained by  modifying the mixing matrix. In such a case one obtain a constrain on the singlet-octet mixing and its $s$-quark content.

\item The contribution of the ${\cal O}_{11}$ operator  is around 30 \% in decay channels involving  the  $K_0^*$.   The  ${\cal O}_{11}$ contributions  approximately keep the  $SU(3)$ relations between different decay  channels.

\item The chiral enhancement predicted by  the NFA  could be used to test the quark structure of the heavy multiplet. Strong deviations from the NFA results
could be interpreted as a signal that the heavy scalars are not  pure two quark state.
\end{itemize}

\section{Aknowledgement}
\noindent CONACyT support under contracts  46195 and 57970 as well
as PROMEP support is gratefully acknowledged. C.R. and J.M. want
to thank the Physics Institute of Guanajuato University for their
hospitality.

\appendix*

\section{NAIVE FACTORIZATION APPROACH (NFA)}

\noindent The relevant  effective Hamiltonian  is given by  \cite{nfa} :

\begin{eqnarray}
{\cal H}_{\text eff} &=& {G_F \over \sqrt{2}} \, \left[ V_{ub}
V_{uq}^* \left (C_1 O_1^u + C_2 O_2^u \right) - V_{tb} V_{tq}^* \,
\left(\sum_{i=3}^{10} C_i \, O_i + C_g O_g  \right) \right] +
\text{h.c.} \label{eq:hamil}
\end{eqnarray}

\noindent where $\lambda_{q^\prime q} = V_{q^\prime b} V^*_{q^\prime q}$, with
$q=d,s$, while $q^\prime = u,c,t$. The Cabibbo-Kobayashi-Maskawa
(CKM) matrix elements are denoted by $ V_{ij}$.  $O_i$ stand for the following four fermion operators:

\begin{eqnarray}
&{\cal O}_1  = (\bar q u)_L(\bar u b)_L,   \hskip2cm & {\cal O}_2 = (\bar u_\alpha b_\beta)_L(\bar q_\beta u_\alpha)_L,  \nonumber \\
&{\cal O}_3  = (\bar q b)_L \sum_{q'}(\bar q'q')_L,  &{\cal O}_4 =(\bar q_\alpha b_\beta)_L \sum_{q'}(\bar q_\beta' q_\alpha')_L, \nonumber \\
&{\cal O}_5= (\bar q b)_L \sum_{q'}(\bar q'q')_R, &
  {\cal O}_6 =(\bar q_\alpha b_\beta)_L \sum_{q'}(\bar q_\beta'q_\alpha')_R=-2\sum_{q'}(\bar q'b)_{S-P}(\bar q q')_{S+P}, \nonumber \\
&{\cal O}_7 = {3\over 2}(\bar q b)_L \sum_{q'}e_{q'}(\bar q'q')_R,
& {\cal O}_8 = {3\over 2}(\bar q_\alpha b_\beta)_L\sum_{q'}
e_{q'}(\bar q_\beta'q_\alpha')_R= -3\sum_{q'}e_{q'}(\bar q'b)_{S-P}(\bar q q')_{S+P}  , \nonumber  \\
&{\cal O}_9  = {3\over 2}(\bar q b)_L \sum_{q'}e_{q'}(\bar q'
q')_L,  & {\cal O}_{10} = {3\over 2}(\bar q_\alpha b_\beta)_L
\sum_{q'} e_{q'}(\bar q_\beta' q_\alpha')_L = {3\over 2} \sum_{q'}
e_{q'}(\bar q'b)_L (\bar qq')_L.
\end{eqnarray}

\noindent and the dipole penguin operator:

\begin{eqnarray*}
{\cal O}_{11} = {g_s \over 16 \pi^2} m_b \bar q \sigma_{\mu \nu} R
T_a b G_a^{\mu \nu};
\end{eqnarray*}

\noindent with $T_a=\lambda_a/2$ the $SU(3)_C$ generators. The Wilson
coefficients  $C_i$ appear in the combinations
$a_{2i-1} = C_{2i-1} + C_{2i}/N$, $a_{2i} = C_{2i} + C_{2i-1}/N$.
The numerical values are taken from \cite{nfa}. Similarly we
define $a_{11} = (8/9)\alpha_s C_{11} (m_b^2/4 \pi q^2)\simeq
-5.7\cdot 10^{-3}$. For the gluon momentum we use  $q\simeq
p_b-p_s\simeq p_B-p_k/2$, so $q^2\simeq m_B^2/2$ \cite{O11}.
Taking $\alpha_s(q^2\simeq m_B^2/2)=0.21$ and $C_{11}=-0.29$ one
obtains $a_{11}\simeq -5.7\cdot 10^{-3}$. Chiral projections are
$L,\ R=1\mp \gamma_5$. Using the relation $2(T_i)_{\alpha \beta} (T_i)_{\gamma \delta} = \delta_{\alpha
\delta} \delta_{\beta \gamma} - (1/N) \delta_{\alpha \gamma}
\delta_{\beta \delta} $ and the Fiertz reordering one obtains for ${\cal O}_{11}$ \cite{O11}
\bigskip

\begin{eqnarray}
{\cal H}_{11} &=& i{G_F\over \sqrt{2}}\lambda_{tq} {C_g\alpha_Sm_b\over 8\pi k^2} \left[{N_C^2-1\over N_C^2}\delta_{\alpha \beta} \delta_{\gamma \delta}-{2\over N_C}T^a_{\alpha \beta}T^a_{\gamma \delta} \right]\nonumber\\
&& k_\mu\left[3i\bar q_\alpha R\gamma^\mu  q'_\beta \bar q'_\gamma R b_\delta
-3i\bar q_\alpha R q'_\beta \bar q'_\gamma \gamma^\mu R b_\delta
+\bar q_\alpha R\gamma_\nu q'_\beta \bar q'_\gamma \sigma^{\nu\mu} R b_\delta -
\bar q_\alpha R \sigma^{\mu\nu} q'_\beta \bar q'_\gamma  \gamma_\nu R b_\delta \right]
\end{eqnarray}

\noindent where $k^2\simeq m_B/2-m_K^2/8$.

\noindent The amplitudes, including  $O_{11}$ contribution, in the NFA
are given by:

\begin{eqnarray*}
A_{\bar B^0\to \pi^-a_0^+}&\simeq&
\lambda_{ud}(a_1X^{\pi^-}_{\bar B^0 a_0^+}+a_2X^{\bar B^0}_{(a_0^+\pi^-)_u})-\lambda_{td}\Biggl[ \left(a_4+a_{10}-{(a_6+a_8)m_\pi^2\over \hat m (m_b+m_u)}\right)X^{\pi^-}_{\bar B^0a_0^+}\nonumber \\
&& +\left(2(a_3-a_5)+a_4+{a_9-a_7-a_{10}\over 2}-{(a_6-a_8/2)m_B^2\over m_u(m_b+m_d)}\right)X^{\bar B^0}_{(a_0^+\pi^-)_u} \Biggr] \nonumber \\
A_{\bar B^0\to \pi^+a_0^-}&\simeq&
\lambda_{ud}(a_1X^{a_0^-}_{\bar B^0 \pi^+}+a_2X^{\bar B^0}_{(a_0^-\pi^+)_u})-\lambda_{td}\Biggl[ (a_4+a_{10})X^{a_0^-}_{\bar B^0\pi^+}-2(a_6+a_8)\tilde X^{a_0^-}_{\bar B^0\pi^+}\nonumber \\
&& +\left(2(a_3-a_5)+a_4+{a_9-a_7-a_{10}\over 2}-{(a_6-a_8/2)m_B^2\over m_u (m_b+m_d)}\right)X^{\bar B^0}_{(a_0^-\pi^+)_u} \Biggr]
\end{eqnarray*}

\begin{eqnarray*}
A_{B^-\to \pi^-S^0}&\simeq&
\lambda_{ud}a_1(X^{\pi^-}_{B^-S^0}+X^{B^-}_{S^0\pi^-})-\lambda_{td}\Biggl[ \left(a_4+a_{10}-{(a_6+a_8)m_\pi^2\over \hat m (m_b+m_u)}\right)X^{\pi^-}_{B^-S^0}\nonumber \\
&& +\left(a_4+a_{10}-{(a_6+a_8)m_B^2\over \hat m (m_b+m_u)}\right)X^{B^-}_{S^0\pi^-}+(a_8-2a_6)\tilde X^{S^0_d}_{B^-\pi^-} \Biggr] \nonumber \\
A_{B^-\to \pi^0a_0^-}&\simeq&
\lambda_{ud}\left[a_1(X^{a_0^-}_{B^- \pi^0}+X^{B^-}_{a_0^-\pi^0})+a_2X^{\pi^0_u}_{B^-a_0^-}\right]-\lambda_{td}\Biggl[ \left(a_4+a_{10}\right)X^{a_0^-}_{B^-\pi^0}-2(a_6+a_8)\tilde X^{a_0^-}_{B^-\pi^0}\nonumber \\
&&- \left(a_4-{3\over 2}(a_9-a_7)-{1\over 2}a_{10}-{(a_6+a_8)m_\pi^2\over m_u(m_b+m_d)} \right)X^{\pi_u^0}_{B^-a_0^-}+\left(a_4+a_{10}-{(a_6+a_8)m_B^2\over \hat m(m_b+m_u)}\right)X^{B^-}_{a_0^-\pi^0} \Biggr] \nonumber \\
A_{\bar B^0\to \pi^0S^0}
&\simeq& \lambda_{ud}a_2(X^{\pi^0_u}_{\bar B^0a_0^0}+X^{\bar B^0}_{(a_0^0\pi^0)_u})-\lambda_{td}\Biggl[ \left(-{3\over 2}a_7+{(2a_6-a_8)m_\pi^2\over 2m_d(m_b-m_d)}\right)X^{\pi^0_u}_{\bar B^0a_0^0}\nonumber \\
&& +\left(2a_5+{3\over 2}a_7+{(2a_6-a_8)m_B^2\over 2m_d(m_b+m_d)}\right)X^{\bar B^0}_{(a_0^0\pi^0)_u}\Biggr]
\end{eqnarray*}

\begin{eqnarray*}
A_{\bar B^0\to K^-a_0^+}&\simeq&
\lambda_{us}a_1X^{K^-}_{\bar B^0\pi^+}-\lambda_{ts}\Biggl[\Bigl(a_4+a_{10}-(a_6+a_8)r_\chi^K \Bigr)X^{K^-}_{\bar B^0a^+}\nonumber \\
&&+\left(a_4-{1\over 2}a_{10}-{(2a_6-a_8)m_B^2\over (m_b+m_d)(m_s+m_d)}\right)X^{\bar B^0}_{K^-a_0^+} \Biggr]  \nonumber \\
A_{B^-\to K^-S^0}&\simeq&
\lambda_{us}a_1\left[X^{K^-}_{B^-S^0}+X^{B^-}_{S^0K^-}\right]
-\lambda_{ts}\Biggl[\left(a_4+a_{10}-(a_6^{\rm eff.}+a_8)r_\chi^K\right) X^{K^-}_{B^-S^0}-\left(2a_6-a_8-{30\over 32}a_{11}\right)\tilde
X^{S_s^0}_{B^-K^-} \nonumber \\
&&+\left(a_4+a_{10}-{2(a_6+a_8)m_B^2\over(m_b+m_u)(m_s+m_u)}\right)X^{B^-}_{S^0K^-}\Biggr] \nonumber \\
A_{B^-\to \bar K^0a_0^-}&\simeq&
\lambda_{us}a_1X^{B^-}_{\bar K^0a_0^-}-\lambda_{ts}\Biggl[\Bigl(a_4-{1\over 2}a_{10}-(a_6-a_8/2)r_\chi^K \Bigr)X^{\bar K^0}_{B^-a_0^-}\nonumber \\
&& +\left(a_4+a_{10}-{2(a_6+a_8)m_B^2\over(m_b+m_u)(m_s+m_u)}\right)X^{B^-}_{\bar K^0a_0^-}\Biggr]
\end{eqnarray*}

\begin{eqnarray*}
A_{\bar B^0\to \bar K^0S^0}&\simeq&
-\lambda_{ts}\Biggl[\left(a_4-{a_{10}\over 2}-(a_6^{\rm eff.}-a_8/2)r_\chi^K\right) X^{\bar K^0}_{\bar B^0S^0}-\left(2a_6-a_8-{30\over 32}a_{11}\right)\tilde X^{S_s^0}_{\bar B^0\bar K^0}\nonumber \\
&&+\left(a_4-{a_{10}\over 2}-{(2a_6-a_8)m_B^2\over(m_b+m_d)(m_s+m_d)}\right)X^{\bar B^0}_{S^0\bar K^0}\Biggr]
\end{eqnarray*}

\begin{eqnarray*}
A_{\bar B^0\to \pi^+{K_0^*}^-}&\simeq&
\lambda_{us}a_1X^{{K_0^*}^-}_{\bar B^0\pi^+}-\lambda_{ts}\Biggl[\Bigl(a_4+a_{10}-(a_6^{\rm eff.}+a_8)r_\chi^*\Bigr)X^{{K_0^*}^-}_{\bar B^0\pi^+}\nonumber \\
&&+\left(a_4-{1\over 2}a_{10}-{(2a_6-a_8)m_B^2\over (m_b+m_d)(m_s+m_d)}\right)X^{\bar B^0}_{\pi^+{K_0^*}^-} \Biggr]  \nonumber \\
A_{B^-\to \pi^- \bar {K_0^*}^0}&\simeq&
\lambda_{us}a_1X^{B^-}_{\bar {K_0^*}^0\pi^-}-\lambda_{ts}
\Biggl[\Bigl(a_4-{1\over 2}a_{10}-(a_6^{\rm eff.}-a_8/2)r_\chi^*\Bigr)X^{\bar {K_0^*}^0}_{B^-\pi^-}\nonumber \\
&& +\left(a_4+a_{10}-{2(a_6+a_8)m_B^2\over(m_b+m_u)(m_s+m_u)}\right)X^{B^-}_{\bar {K_0^*}^0\pi^-}\Biggr]\nonumber \\
A_{B^-\to \pi^0{K_0^*}^-}&\simeq&
\lambda_{us}\left[a_1\left(X^{{K_0^*}^-}_{B^-\pi^0}+X^{B^-}_{\pi^0 {K_0^*}^-}\right)+a_2 X^{\pi^0_u}_{B^-{K_0^*}^-}\right]
-\lambda_{ts}\Biggl[\left(a_4+a_{10}-(a_6^{\rm eff.}+a_8)r_\chi^* \right) X^{{K_0^*}^-}_{B^-\pi^0}\nonumber \\
&&+\left(a_4+a_{10}-{2m_B^2(a_6+a_8)\over(m_b+m_u)(m_s+m_u)}\right) X^{B^-}_{{K_0^*}^-\eta}
+{3\over 2}(a_9-a_7)X^{\pi^0_u}_{B^-{K_0^*}^-}\biggr] \nonumber \\
A_{B^-\to \eta{K_0^*}^-}&\simeq&
\lambda_{us}\left[a_1\left(X^{{K_0^*}^-}_{B^-\eta}+X^{B^-}_{\eta {K_0^*}^-}\right)+a_2 X^{\eta_u}_{B^-{K_0^*}^-}\right]
-\lambda_{ts}\Biggl[\left(a_4+a_{10}-(a_6^{\rm eff.}+a_8)r_\chi^* \right) X^{{K_0^*}^-}_{B^-\eta}\nonumber \\
&&
+\left({3\over 2}(a_9-a_7)-2a_4+a_{10}+(2a_6-a_8)r_\chi^{\eta_s}+{a_{11}\over 64}\left(12-30r_\chi^{\eta_s}-1\right)\right)X^{\eta_u}_{B^-{K_0^*}^-}\biggr]\nonumber \\
&&+\left(a_4+a_{10}-{2m_B^2(a_6+a_8)\over(m_b+m_u)(m_s+m_u)}\right) X^{B^-}_{{K_0^*}^-\eta} \nonumber \\
\end{eqnarray*}

\begin{eqnarray}
A_{\bar B^0\to \pi^0{\bar{K_0^*}}^0}&\simeq&
\lambda_{us}a_2X^{\pi^0}_{\bar B^0{\bar{K_0^*}}^0}-\lambda_{ts}\Biggl[\Bigl(a_4-{a_{10}\over 2}-(a_6^{\rm eff.}-a_8/8)r_\chi^*\Bigr)X^{\bar{{K_0^*}}^0}_{\bar B^0\pi^0}+
{3\over 2}(a_9-a_7)X^{\pi^0}_{\bar B^0 {\bar{K_0^*}}^0} \nonumber \\
&&+\left(a_4-{a_{10}\over 2}-{(2a_6-a_8)m_B^2\over (m_b+m_d)(m_s+m_d)}\right)X^{\bar B^0}_{{\bar {K_0^*}}^0\pi^0} \Biggr] \nonumber \\
A_{\bar B^0\to \eta{\bar{K_0^*}}^0}&\simeq&
\lambda_{us}a_2X^{\eta_u}_{\bar B^0{\bar{K_0^*}}^0}-\lambda_{ts}\Biggl[\Bigl(a_4-{a_{10}\over 2}-(a_6^{\rm eff.}-a_8/2)r_\chi^*\Bigr)X^{\bar{{K_0^*}}^0}_{\bar B^0\eta}\nonumber \\
&&+\left({3\over 2}(a_9-a_7)-2a_4+a_{10}+(2a_6-a_8)r_\chi^{\eta_s}+{a_{11}\over 64}\left(12-30r_\chi^{\eta_s}-1\right)\right) X^{\eta_u}_{\bar B^0 {\bar{K_0^*}}^0}\nonumber \\
&&+\left(a_4-{a_{10}\over 2}-{(2a_6-a_8)m_B^2\over (m_b+m_d)(m_s+m_d)}\right)X^{\bar B^0}_{{\bar {K_0^*}}^0\eta} \Biggr]
\end{eqnarray}

\noindent and $A_{\bar B^0\to {K_0^*}^+\pi^-}=0$, $S^0=a_0^0,\ \sigma$ and $f_0$, $r_\chi^*\simeq 2m_{K_0^*}^2/m_bm_s$, $r_\chi^{\eta_s}\simeq m_\eta^2/m_bm_s$ and $a_6^{\rm eff.}r_\chi^M=a_6r_\chi^M-a_{11}\left[12(1-r_\chi^M)-1\right]/32$.

\begin{eqnarray}
&&X^{K^-}_{B^-f_0} =<K^-|(\bar us)_L|0><f_0|(\bar
ub)_L|B^->=f_K(m_B^2-m_{f_0}^2)F_0^{B^-f_0}(m_\pi^2)=f_K{F_0^{\bar
B^0a_0^+}(m_K^2)\over\sqrt{2}}(m_B^2-m_{f_0}^2)\sin\phi_S
\nonumber \\
&&X^{\bar K^0}_{\bar B^0f_0} =<\bar K^0|(\bar sd)_L|0><f_0|(\bar
db)_L|\bar B^0>=f_{\bar K^0}(m_B^2-m_{f_0}^2)F_0^{\bar
B^0f_0}(m_K^2)=f_K{F_0^{\bar B^0a_0^+}(m_K^2)\over\sqrt{2}}
(m_B^2-m_{f_0}^2)\sin\phi_S\nonumber \\
&&X^{K^{*-}_0}_{\bar B^0 \pi^+} =<K^{*-}_0|(\bar su)_L|0><\pi^+|(\bar ub)_L|\bar B^0>=f_{K^{*-}_0}(m_B^2-m_\pi^2)F_0^{\bar B^0\pi^+}(m_{K_0^*}^2)=-f_{K^{*-}_0}(m_B^2-m_\pi^2)F_0^{ B^0\pi^-}(m_{K_0^*}^2)\nonumber \\
&&X^{\bar K^{*0}_0}_{B^- \pi^-} =<\bar K^{*0}_0|(\bar sd)_L|0><\pi^-|(\bar db)_L|B^->=f_{\bar K^{*0}_0}(m_B^2-m_\pi^2)F_0^{\bar B^-\pi^-}(m_{K_0^*}^2)=-f_{K^{*0}_0}(m_B^2-m_\pi^2)F_0^{ B^0\pi^-}(m_{K_0^*}^2)\nonumber \\
&&X^{K^{*-}_0}_{B^- \pi^0} =<K^{*-}_0|(\bar su)_L|0><\pi_0|(\bar
ub)_L|B^->=f_{K^{*-}_0}(m_B^2-m_{\pi^0}^2)F_0^{B^-\pi_0}(m_{K_0^*}^2)
=-f_{K_0^*}{F_0^{B^0\pi^-}(m_*^2)\over\sqrt{2}}\nonumber \\
&&X^{\pi^0_u}_{\bar B^0 {\bar K^{*0}}_0}=<\pi^0|(\bar uu)_L|0><{\bar K^{*0}}_0|(\bar sb)_L|\bar B^0>={f_\pi\over \sqrt{2}}(m_B^2-m_{K_0^*}^2)F_0^{\bar B^0{{\bar K^{*0}}_0}}(m_\pi^2)={f_\pi\over \sqrt{2}}(m_B^2-m_{K_0^*}^2)r_{K\pi}F_0^{\bar B^0a_0^+}(m_\pi^2)\nonumber \\
&&X^{{\bar K^{*0}}_0}_{\bar B^0 \pi^0}= <K^{*0}_0|(\bar
sd)_L|0><\pi^0 |(\bar db)_L|\bar
B^0>=f_{K^{*0}_0}(m_B^2-m_{\pi^0}^2)F_0^{\bar
B^0\pi^0}(m_{K_0^*}^2)=f_{K_0^*}\left(m_B^2-m_\pi^2\right){F_0^{B^0\pi^-}(m_*^2)\over\sqrt{2}} \nonumber \\
&&\tilde X^{{f_0}_d}_{B^-\pi^-}= <f_0|\bar dd|0><\pi^-|\bar d
b|B^->=m_S\bar f_{f_0}^d{m_B^2-m_\pi^2\over
m_b-m_d}F_0^{B^-\pi^-}(m_{f_0}^2)=-{1\over
\sqrt{2}}{m_B^2-m_\pi^2\over m_b-m_d}m_{f_0}\bar
f_{f_0}^nF_0^{B^0\pi^-}(m_{f_0}^2) \nonumber \\
&&\tilde X^{{f_0}_s}_{B^-K^-} = <f_0|\bar ss|0><K^-|\bar s
b|B^->=m_{f_0}\bar f_{f_0}^s{m_B^2-m_K^2\over
m_b-m_s}F_0^{B^-K^-}(m_{f_0}^2)={m_B^2-m_K^2\over
m_b-m_s}r_{K\pi}m_{f_0}\bar f_{f_0}^sF_0^{B^0\bar\pi^-}(m_{f_0}^2)\nonumber \\
&&\tilde X^{{f_0}_s}_{\bar B^0\bar K^0} = <f_0|\bar ss|0><\bar
K^0|\bar s b|\bar B^0>=m_{f_0}\bar f_{f_0}^s{m_B^2-m_K^2\over
m_b-m_s}F_0^{\bar B^0\bar K^0}(m_{f_0}^2)=\tilde
X^{{f_0}_s}_{B^-K^-}
\end{eqnarray}

\noindent for annihilation:

\begin{eqnarray}
&&X^{B^-}_{f_0K^-} = <f^0K^-|(\bar su)_L|0><0|(\bar u b)_L|B^->=-f_B(m_S^2-m_K^2)F_0^{f^0K^-}(m_B^2)\nonumber \\
&&X^{\bar B^0}_{\bar K^0f_0} =<\bar K^0f^0|(\bar sd)_L|0><0|(\bar
d b)_L|\bar B^0>=-f_B(m_{f_0}^2-m_K^2)F_0^{f_0\bar K^0}(m_B^2)
=X^{B^-}_{K^-f_0} \nonumber \\
&&X^{\bar B^0}_{K^{*-}_0\pi^+} =<K^{*-}_0\pi^+|(\bar sd)_L|0><0|(\bar d b)_L|\bar B^0>=-f_B(m_{K^*}^2-m_\pi^2)F_0^{K^{*-}_0\pi^+}(m_B^2)\nonumber \\
&&X^{B^-}_{\bar K^{*0}_0\pi^-} =<\bar K^{*0}_0\pi^-|(\bar su)_L|0><0|(\bar u b)_L|B^->=-f_B(m_{K^*}^2-m_\pi^2)F_0^{\bar K^{*0}_0\pi^-}(m_B^2)\nonumber \\
&&X^{B^-}_{K^{*-}_0\pi^0} =  <K^{*-}_0\pi^0|(\bar su)_L|0><0|(\bar u b)_L|B^->=-f_B(m_{K^*}^2-m_{\pi^0}^2)F_0^{K^{*-}_0\pi^0}(m_B^2)  \nonumber \\
&&X^{\bar B^0}_{{\bar K^{*0}}_0\pi^0} =<{\bar K^{*0}}_0\pi^0|(\bar
sd)_L|0><0|(\bar d b)_L|\bar
B^0>=-f_B(m_{K^*}^2-m_\pi^2)F_0^{{\bar K^{*0}}_0\pi^0}(m_B^2)
\end{eqnarray}

\noindent with $r_{K\pi}=F^{BK}/F^{B\pi}\simeq f_K/f_\pi\simeq 1.21(9)$

\end{document}